\documentclass[12pt]{amsart}

\usepackage{epsfig}
\usepackage{xcolor}
\usepackage{hyperref} 
\usepackage[utf8]{inputenc}
\usepackage[russian,english]{babel}

\def\h{\hbar}
\def\CC{\mathbb C}
\def\RR{\mathbb R}

\def\H{\mathcal H}
\def\bra{\langle}
\def\ket{\rangle}
\def\tr{\operatorname{tr}}
\def\P{\mathcal P}

\def\U{\mathcal U}
\def\hat{\widehat}

\begin{document}
\title{Intelligent Qubits}
\author[Alexander Givental]{Alexander Givental}
\address{Department of Mathematics, University of California Berkeley}
\email{sumizdat@berkeley.edu}
\begin{abstract} We introduce the method of ``intelligent qubits,'' which replace live observers in Wigner's friend and Frauchiger--Renner thought experiements, in order to expose the source of the paradoxes: tacit substitution of Boolean hidden variables for noncommuting quantum propositions. The fallacy traces back to London and Bauer's attempt (1939) to legitimize a self-referential interpretation of observers' ``introspection,'' and thus intertwines the measurement and mind-body problems. We cut the Gordian knot by abandoning reductionism, and describe an alternative which fits, we argue, equally naturally with both physics and anthropology takes on the problems. \end{abstract}

\maketitle

{\bf 0. Introduction.} The scope of this paper is twofold. Our narrow target is the measurement problem in quantum mechanics, which belongs to methodology of physics. But it leads us to the aspects of that methodology which could be considered anthropological and pertain to philosophy.

The latter, for its bi-millennial lifespan, has earned in scientific circles the reputation \cite{Aha} of a field that prides itself on {\em discussing} fundamental problems, such as e.g. ``freedom of will'' or ``the hard problem of consciousness,'' without ever resolving any --- or even trying to.

The research area of {\em quantum interpretations}, though only a hundred years old \cite{QI} has come to a state somewhat resembling that of philosophy. Specifically, current research dedicated to the measurement problem in quantum mechanics, which was characterized in a recent survey article \cite{TMB} as a ``forest'' and ``the cacophony of competing interpretations'', is distributed among hundreds of publications featuring mind-bending thought experiments, defying common sense ``no-go'' theorems, and subtle new concepts needed to distinguish various approaches. The problem with this is not that the subject requires a 60-page review --- this is normal or even modest for any active field --- but that the many ``leading'' interpretations are {\em competing}, which implies that they are hardly compatible with each other, yet none of them provides satisfactory answers.

Planting one more tree and see how it fares compared to the rest of the forest is not exactly our intention. Rather, we aim to show that the mainstream quantum mechanics as it was understood by its classics fits naturally into a simple and reaistic worlview, whereas it is the traditional physicalist paradigm of an objectively functioning dynamical universe that emerges as counterfactual. For this, in the first part, we deconstruct the measurement problem to see that --- paradoxically from the physicalist perspective --- the ``cut'' between the object and the subject is impenetrable, and in the second part, we find that an anthropological outlook leads to exact same conclusion.

More specifically: In Section 1, we review the tensions associated with the quantum measurement problem to show that they are already present in classical mechanics, and point out key differences. In Section 2, we recall the von Neumann measurement model, stressing that it does not reduce irreversible measurement to reversible dynamics. In Section 3, we uncover the irreducible role of dissipative memory devices, and thereby answer the question of what an act of measurement is. Here we also frame quantum measurement as a transformation of quantum propositions to propositions of Boolean logic. In Section 4, we show that quantum measurements are necessarily inter-subjective, and introduce the notion of the {\em universal observer.} This suggests that treating an observer as a part of an isolated quantum system should lead to Russell's self-reference paradoxes. Besides, measurement outcomes presumably accessible to such an observer are akin to ``hidden variables'' disallowed by Bell's inequalities, an instance of which is proved in Section 5. In Section 6 we show how both Russell's and Bell's conflicts manifest in Wigner's friend thought experiment and that the ambiguity of interpretation in it is a result of tacitly treating non-commuting quantum propositions as commuting Boolean variables. The fog around the mind--body problem can be blamed for this confusion. To escape the trap and illuminate the logical structure of the situation, we introduce {\em intelligent qubits}: the observers' minds whose binary states are to be measured in unrealistic thought experiments are replaced with perfectly measurable qubits whose only unrealistic property is that they can reason, apply quantum mechanics, and exchange information. The method is applied once again in Section 7 where we deconstruct Frauchiger--Renner's generalization \cite{FR} of Wigner's friend setup and expose the same interrelated fallacies. All this motivates a worldview, laid out in Section 8, where the universal observer and the {\em black box} of its environment are clearly separated and coexist on an equal footing. This worldview also points toward a specific resolution of the fireball paradox in the theory of Hawking radiation. In Section 9, which opens the anthropological part of the paper, we give a working definition of {\em culture}, which is nominated to the role of the universal observer. In Section 10, the ``objective reality'' turns out to be the history of interaction between the universal observer and the black box. Section 11 presents an argument which, we hope, blows the mind--body fog away. Section 12 explains why epistemic nature of quantum states doesn't undermine their ontic status. In Section 13, the Berkleyan aspect of quantum mechanics that ``things don't exist unless observed'' is reconciled with our macroscopic classical experience. In Section 14, we review the usual physics of dissipative memory media to see how it squares with their irreducible role in measurement proclaimed in Section 3.





We should add that this paper differs from two other available expositions: from \cite{GivQ} by being more substantial and less mathematical, and from \cite{GivM} (which was rather popularizing than scientific as it didn't even contain any bibliography) by expecting the reader to be familiar with basic quantum mechanical notation. Still, even here the second part is almost entirely nontechnical.   

\section*{Physics}

{\bf 1. The quantum measurement problem.} A pure state of an {\em isolated} quantum mechanical system is represented by a wave function which is a unit vector $|\psi\ket$ (or simply $\psi$) in a Hilbert vector space $\H$. Time evolution of the state is deterministic and is abstractly described by the family $|\psi\ket \mapsto e^{tH/i\h}|\psi\ket$ of unitary transformations. It obeys the Schr\"odinger equation $i\h \dot{\psi} = H \psi$ which is the quantum analogue of a classical Hamiltonian mechanical system.

Yet, the vector $\psi \in \H$ remains detached from any physical reality unless one engages in measuring physical quantities --- observables. Such observables are mathematically represented by Hermitian operators $A: \H\to \H$, and the one and only role of $\psi$ is to specify the {\em expectation value}
of the outcome for every observable $A$ in accordance with {\em Born's rule}.  That is, outcomes of individual measurements are genuinely, inherently random, and as Nobel-winning experiments concerning Bell's inequalities show, this randomness is not due to our incomplete knowledge. Nonetheless, if one engages in a series of identical measurements of the quantity impersonated by $A$ on an ensemble of identical quantum systems prepared to be at the same state $\psi$, the random outcomes will average to $\bra\psi|A|\psi\ket$.

Furthermore, if an act of measurement doesn't destroy the system, it continues to evolve deterministically, and one may ask what initial post-measurement state should be attributed to it. In other words, if right after measuring $A$ one engages in measuring another observable $B$, what is the expectation value of $B$ {\em conditioned} on the outcome of measuring $A$? The answer to this question is given by L\"uders -- von Neumann's {\em projection rule}. The expectation value of $B$ under the condition that the measured value of $A$ was equal to $\lambda$ is $\bra\hat{\psi}|B|\hat{\psi}\ket$, where $\hat{\psi}$ is the normalized orthogonal projection of $\psi$ to the eigenspace of $A$ with the eigenvalue $\lambda$.

Thus, an ideal interaction of a formerly isolated system with a measuring device interrupts the unitary (and hence potentially reversible) evolution of the pure quantum state with what is colloquially known as ``collapse of the wave function''. It consists of two phases, each irreversible: first $\psi$ turns into a statistical mixture of the states $\hat{\psi}$, and then --- upon examining the outcome of the measurement --- into one of them.

The measurement problem is sometimes framed as the conceptual tension between unitary evolution and collapse. Classical textbooks (e.g. \cite{LL}) describe measurement as involving an (often microscopic) quantum system interacting with a macroscopic classical apparatus. But there is no clear boundary (Heisenberg's ``cut'') between microscopic and macroscopic, and the apparatus and even the observer reading the outcome are physical objects, and should be in principle describable, together with the very process of measurement, by the unitary formalism of quantum mechanics (shouldn't they?) The fact that the state $\hat{\psi}$ characterizes expectation values conditioned on the observer's knowledge of the outcomes adds to the confusion, since one now has to explain who (or what?) counts as an observer, decide whether the status of quantum states is merely epistemic, and then figure out what, if anything, represents the actual state of the universe.      

Perhaps this description of the problem is not wrong, but it is somewhat superficial, because the same issues arise in classical mechanics as well. Whereas the dynamics of a classical mechanical system is described by its Hamiltonian trajectory in a symplectic phase space, the initial phase point of a truly isolated system is fundamentally {\em unknowable}. This simple observations leads one into the realm of {\em statistical mechanics}. The state of a classical mechanical system is characterized by a probability density $\rho$ with respect to the Liouville measure $\Omega$ on the phase space $M$. Its evolution is governed by the Poisson bracket with the Hamilton function: $\dot{\rho}=\{ \rho, H_{classical}\}$.  (The quantum analogue of $\rho$ is not the wave function but a density matrix $P$ describing mixed quantum states and subject to the Heisenberg equation $i\hbar \dot{P}=[H,P]$.) Expectation values of classical observables $f: M\to \RR$ are given by the integral $\int_M f\rho\,\Omega$ (its quantum analogue is $\tr AP$). Measurements interrupt the deterministic evolution of the density and ``collapse'' it to $\rho_{new}$ reflecting information gained from the outcome on the location of the phase point (e.g. that the value of $f$ lies in a certain interval, see \cite{GivQ}). The same methodological questions one can ask about the quantum collapse apply to the classical one. 

One reason why these questions are ignored in classical physics is that the latter originated in celestial mechanics where one could not make observations intrusive even if one wanted to. Another reason is that, theoretically speaking, one can use successive observations together with precise description of interim dynamics in order to localize (``collapse'') the density $\rho$ to a vicinity of a single phase point. (In practice this doesn't work due to dynamical chaos and non-integrability.)

In any case, excluding observers from classical mechanics doesn't lead to logical contradictions because {\em classical observables commute}. This means that they can, at least in principle, be measured simultaneously, and it is safe to assume that their objective values exist and specify a single point in the phase space (known to God even if not to human observers). 

In quantum mechanics, where observables don't commute, it doesn't work, and successive measurements of non-commuting observables can erase information gained from prior measurements rather than augment it. But the most perplexing side of quantum measurement is that the decisions of what to measure and in what order is made by the observers, who therefore inevitably influence the fate of the system. The need to resolve the impasse between such active role of observers and the classical picture of an objectively functioning universe lies at the core of the measurement paradox.

\medskip

{\bf 2. The von Neumann measurement model.} An early attempt of describing measurements via unitary evolution is found in von Neumann's classical monograph \cite{vonN}. One considers an isolated meta-system S$+$A with the Hilbert space $\H\otimes \H'$ where the factors are the Hilbert spaces of the formerly isolated system S and of the measuring apparatus A respectively. One assumes that the initial decomposable state $\psi\otimes \phi$, where $\psi\in \H$ and $\phi\in \H'$ represent the states of S and A before the measurement, is transformed unitarily into an entangled state
\[ \Psi = \sum_\lambda c_\lambda (\psi_\lambda \otimes \phi_\lambda). \]
Here $\phi_\lambda \in \H'$ is the state of the apparatus representing the value $\lambda$ of the measured observable, $\psi_\lambda\in \H$ is an eigenvector of the observable with the eigenvalue $\lambda$ to which the state $\psi$ of S collapses when $\lambda$ is the outcome, and $c_\lambda\in \CC$ is the complex amplitude whose squared absolute value $|c_\lambda|^2$ represents Born's probability of the outcome $\lambda$ (i.e. it is a Fourier coefficient of the state $\psi$):
\[ \psi = \sum_\lambda c_\lambda \psi_\lambda.\]

In the literature, one can encounter claims \cite{SYL} that the starting point of the measurement problem is a conflict between this unitary model and the collapse description of measurement. In fact the von Neumann model does not yet describe any outcome of the measurement. Namely, the state $\Psi$ of the meta-system is a gadget that specifies expectation values of all observables on $\H\otimes \H'$. To obtain a specific outcome of the measurement, someone outside the meta-system still has to read the state of the apparatus. It is a measurement on S$+$A of the observable $I\otimes A'$, where $I$ is the identity operator on $\H$, and $A':\H'\to \H'$ corresponds to the observable $A:\H\to H$ the apparatus was meant to measure. Of course, one can extend the meta-system to include that someone as a quantum system with a Hilbert space $\H''$, but this would result in an entangled meta-meta-state on which someone else would have to perform a measurement, etc., leading only to an infinite recursion.    

On the other hand, the entangled state $\Psi$ symbolizes a perfect correlation between the states of the apparatus and the system. Namely, the observables $A\otimes I'$ and $I\otimes A'$ commute and can be measured simultaneously, which with probability one will result in the same value of $\lambda$. There is an important subtlety here, and so let's spell it out.

In the original collapse picture of measuring the observable $A$ on the system S, repeated measurements of $A$ (which commutes with itself) yield the same value $\lambda$ with probability one. Indeed, the projection rule results in the collapsed state $\psi_\lambda$ which is an eigenvector of $A$ orthogonal to all eigenvectors $\psi_\mu$ with $\mu\neq \lambda$, leaving zero chances of any other than $\lambda$ value of $A$ in subsequent measurements. 

In the von Neumann model, the first measurement of $A$ is recast as reading the value $\lambda$ from the apparatus, i.e. as measuring $I\otimes A'$ on the state $\Psi$. This results in the collapsed state $\psi_\lambda\otimes \phi_\lambda$ of the meta-system. The subsequent measurement of $A$ are now described as measuring $A\otimes I'$ on this collapsed state. Since it is an eigenvector of $A\otimes I'$, the outcome will be the eigenvalue $\lambda$ with probability one.

Returning to the question about the conflict between the two descriptions of measurement, we should note that on the one hand, the von Neumann model cannot lead to any formal contradictions, because entangled states are easily created by experimenters. One often speaks of two entangled spins which ``measure each other'' in the sense that measuring the state of one of them determines the state of the other. On the other hand, the model is used in setups like Wigner's friend and its generalizations \cite{SYL} --- where the ``apparatus'' includes the whole lab, and certain measurement on S+A can only be done by omnipotent super-observers in thought experiments --- in a way that seems to create paradoxes. However, before looking at them, we need to address another, more pressing issue.    

\medskip

{\bf 3. Memory.} The first clue that something is wrong with the classical science paradigm concerns not the role of observers but probabilities. In order to make sense of quantum states one needs to average the outcomes of past observations, which therefore have to be {\em memorized}. The memory media, be it paper, silicon, or brain, act irreversibly by design and purpose. Therefore they involve dissipative processes. Although such processes are usually considered ``secondary'', reducible to fundamental, conservative ones --- their role is not: without them those deemed ``primary'' simply cannot be quantified. In traditional approaches to foundations of physics this issue seems to be suppressed. Below we take a look at a typical attempt to describe the dissipative act in conservative terms of the von Neumann model.

To simplify notations (and prepare for later analysis) note that instead of measuring the values of arbitrary quantum observables $A$ it suffices to check whether the values of $A$ lie in prescribed intervals $(a,b)$ of the number line. The answers $1$ for ``yes'' and $0$ for ``no'' define a proposition of classical Boolean logic. It corresponds to a certain {\em quantum proposition}, namely the orthogonal projector $P$ to the spectral subspace of $A$ corresponding to the interval $(a,b)$. The expectation value $p:=\bra\psi|P|\psi\ket$ is the probability of the event that at the state $\psi$, the value of $A$ lies in this interval. All such probabilities define the distribution of $A$ at the state $\psi$ and are therefore sufficient to determine the expectation value $\bra\psi|A|\psi\ket$. Thus, measuring $\psi$ is abstractly described as an operation assigning classical propositions to quantum ones. This formalizes a view famously expressed by Bohr \cite{Bohr}:
\begin{quote}
[I]t is decisive to recognize that, however far the phenomena transcend the scope of classical physical explanation, the account of all evidence must be expressed in classical terms. The argument is simply that by the word 'experiment' we refer to a situation where we can tell others what we have done and what we have learned and that, therefore, the account of the experimental arrangement and of the results of the observations must be expressed in unambiguous language with suitable application of the terminology of classical physics.
\end{quote}
 
Next, in the von Neumann model for measuring $P$, it suffices to assume that the measuring apparatus is a {\em qubit}, i.e. that its Hilbert space is the standard Hermitian plane $\CC^2$ whose orthonormal basis vectors $|1\ket$, $|0\ket$ are the eigenvectors with the respective eigenvalues of the projector $Q=\left[\begin{array}{cc}1&0\\0&0\end{array}\right]$ (in the role formerly designated to $A'$). The entangled state after the measurement is
therefore
\[ |\Psi\ket=|\psi_1\ket\otimes |1\ket + |\psi_0\ket \otimes |0\ket \in \H\otimes \CC^2\]
where
\[ \psi_1=P\psi, \ \psi_0=(I-P)\psi,\ p=\bra\psi_1|\psi_1\ket=1-\bra\psi_0|\psi_0\ket,\ \bra\psi_1|\psi_0\ket=0.\]

Thus, we have two descriptions of measuring $P$. The direct one yields the answer $1$ or $0$ with probabilities $p$ and $1-p$, a specific outcome is recorded in a memory device (let's call it the lab journal), and the state $\psi$ collapses to one of $\hat{\psi}_i:=\psi_i/|\psi_i|$, $i=1,0$. In the von Neumann model, the system interacts unitarily with the measuring device (the qubit), after which the observer reads the outcome (i.e. measures $I\otimes Q$ on the entangled state), and records the result in the journal. Yet, in order to evade the explicit use of dissipative processes, one brings in one more ingredient --- {\em the environment}, which consists of the rest of the lab together with the observer, her brain, and everything else including the thermal photons which escaped the lab when the observer scratched the surface of the journal. The argument is that the system and the measuring device become unitarily entangled with the uncontrolled environment, after which the environment is discarded, and this discarding accounts for the dissipation. In formal notation, the entangled state is
\[ |\Phi\ket = |\psi_1\ket\otimes |1\ket\otimes |\phi_1\ket +|\psi_0\ket\otimes |0\ket\otimes |\phi_0\ket \in \H\otimes \CC^2\otimes \tilde{\H},\]
where $\phi_i\in \tilde{\H}$ are the states of the environment corresponding to the two outcomes, $\bra\phi_i|\phi_j\ket=\delta_{ij}$. 

The act of ``discarding the environment'' can be described in two equivalent
ways. The first is to say that one intends to use the state $\Phi$ only to measure observables of the form $A\otimes B\otimes \tilde{I}$. The second resorts to the language of density matrices. The entangled pure state $\Phi$ is represented by the rank-one density matrix $|\Phi\ket\bra\Phi|$ (the orthogonal projector to the line of $\Phi$), which after taking the partial trace over $\tilde{\H}$, yields the mixed-state density matrix on $\H\otimes \CC^2$:
\[ \P:=|\psi_1\ket\bra\psi_1|\otimes |1\ket\bra 1| + |\psi_0\ket\bra\psi_0|\otimes |0\ket\bra 0| .\]
The point of this is that the expectation value 
\[ \bra\Phi|A\otimes B\otimes \tilde{I}|\Phi\ket = \tr[(A\otimes B)\P] .\]

The emergence of a mixed state here is an instance of the general phenomenon of {\em decoherence}: using only observables $A\otimes B\otimes \tilde{I}$ results in a loss of information about the pure state $\Phi$ and is equivalent to treating the system S+Q as isolated while in reality it interacts with the environment E.

Yet, the problem with the above construction is that neither the entangled state $\Phi$ nor the density matrix $\P$ specify any outcome. They merely provide an {\em a priori} description of the statistics of possible outcomes (after all, one doesn't need to toss a coin to know it will be 50/50 heads or tails). Namely, when someone from {\em the outside} S+Q+E measures $P\otimes I_2\otimes\tilde{I}$ and $I\otimes Q\otimes \tilde{I}$, the results correlate perfectly:
\begin{align*} \tr[(P\otimes Q)\P) = p, \ \tr[(I-P)\otimes(I_2-Q)\P]&=1-p, \\
    \tr[((I-P)\otimes Q)\P]=\tr[P\otimes(I_2-Q)\P]&=0. \end{align*}
For the actual outcome, an outside observer still needs to run the experiment and then measure $I\otimes Q\otimes \tilde{I}$ (i.e. read and memorize the result). 

Thus, resorting to dissipative processes is only postponed and delegated to the outside observer, but not removed.

\medskip

{\bf 4. Inter-subjectivity and self-reference.} Recall that the end goal of including the environment into the meta-system was to salvage the physicalist (or objectivist) view that everything in the world including dissipative processes has to be described by first principles of fundamental physics. It is clear, however, that this goal is ill-fated anyway, because it ultimately expects that observers can measure themselves --- an assumption that is guaranteed to run into paradoxes of self-reference. We've tacitly avoided this problem by insisting that a system can only be measured from the outside.  

Russell's famous {\em barber's dilemma} uncovers a paradox in naive set theory: the set-defining rule --- ``the barber in a military unit shaves those and only those who don't shave themselves'' --- doesn't define any set. The problem is that the rule applied to the barber himself leads to a contradiction.

Another paradox arises if one introduces the {\em universal set}\  $\U$ which by definition contains all sets --- and hence the set $2^{\U}$ of its own subsets as well. This leads to a contradiction, because the cardinality of $2^{\U}$ is strictly greater than that of $\U$. This is true for any set and means that no function $f: \U\to 2^{\U}$ is surjective. Indeed, assuming that $f$ is surjective, form a subset $F$ by the barber's rule, ``$x\in F \Longleftrightarrow x\notin f(x)$'', and find that $F=f(x_0)$ leads to the same paradox as in barber's dilemma: if $x_0\notin F$ then $x_0\in F$, and if $x_0\in F$, then $x_0\notin F$. 

In set theory, the paradoxes are resolved by the Zermelo--Fraenkel axiomatic approach which stipulates that sets are to be formed in a {\em hereditary} (recursive) fashion following the rules that explicitly evade Russell's self-reference and exclude too big and ambiguously defined ``sets'' like $\U$.

Paradoxes in quantum mechanics, though take the form of statistical discrepancies, also come from applying the machinery to systems which are big enough to include observers and resorting to self-referential reasoning. However, before going into specific examples which would make this intuitively plausible but technically vague observation well-grounded, we need to clarify the notion of the observer by explaining in what sense observations are inter-subjective.

Suppose that preparing for a double-slit experiment and expecting to observe an interference pattern, I remain unaware that my assistant installed a spying device at one of the slits. Then my expectation will disagree with the outcome of the experiment.

This is an illustration to a general rule that in quantum mechanics a pure state
$\psi =\sum_\lambda c_\lambda \psi_\lambda$ before a measurement is not the same as the mixed state after it: the former is characterized by the rank-one density matrix $|\psi\ket\bra\psi|$ and the latter by $\sum_\lambda |c_\lambda|^2|\psi_\lambda\ket\bra\psi_\lambda|$.

Thus, in order to get correct predictions about a system, an observer must be informed about actions on that system of all other observers. He does not necessarily need to know the outcomes of their measurements, but he cannot ignore the fact that such measurements occurred and expect his prediction to be correct.

There is a physical reason for this: a quantum measurement is an idealized but intrusive interaction of a measuring device with the system, and one cannot make reliable predictions by pretending that in the past the system remained isolated when in reality it was not.

In this sense, individual observers are not independent, but form a community -- {\em the universal observer} --- whose members don't have to necessarily share information with each other, but have to be aware of each other's actions on the observed system. As a consequence, one can expect that applying quantum mechanics to an observer (i.e. to someone who is in possession of information about outcomes of some measurements even if this information is not shared with the outside world) would cause statistical inconsistencies, because it qualifies as self-referential reasoning: the universal observer applying quantum mechanics to itself.   

\medskip


{\bf 5. Bell's inequality.} In recent years, a number of imaginary scenarios have been proposed and discussed (see \cite{NR, SYL}) where agents take measurements on and apply quantum mechanics to quantum systems or each other's labs following certain (usually mind-bending) protocols, reason about each other's conclusions, and eventually arrive at conflicting descriptions of the situation, thereby demonstrating incompatibility of various assumptions. The arguments usually combine the premises of Wigner's friend experiment with the logic behind Bell's inequalities.

Finding that one glove in a pair is right implies without further observation that the other one is left, no matter how far from each other the gloves traveled. But in this example of classical entanglement, the gloves retain their rightness and leftness during the travel. In quantum mechanics, entangled qubits don't carry their values until measured. This principle is confirmed by Nobel-winning experiments which show that quantum correlations can exceed the limits which are imposed by Bell's inequalities in the case if the measured properties were predetermined by any ``hidden variables.''

Here is our reconstruction of an argument from \cite{SYL} attributed to L. Hardy and presented as an instance of Bell's inequalities. Consider a system R+S of two qubits and introduce two pairs of quantum propositions measuring each qubit in different bases:  
\begin{align*} R&:=|1\ket\bra 1|\otimes I, \ \ A:=\frac{1}{2}\left(|0\ket-|1\ket\right) \left(\bra 0|-\bra 1|\right) \otimes I, \\
  S&:=I\otimes |1\ket\bra 1\ket, \ \ B:=I\otimes \frac{1}{2}\left(|0\ket-|1\ket\right) \left(\bra 0|-\bra 1|\right). \end{align*}
Note that $R$ and $A$ commute with $S$ and $B$ (though not with each other) and hence can be measured simultaneously.

Consider three implications: $A\Rightarrow S$, $S\Rightarrow R$, $R\Rightarrow \neg B$. Suppose that at some state $\psi$ all three are true with probability one. If the truth values were determined by hidden variables, the Venn diagram in the domain of these variables would consist of three nested sets, i.e. $A\Rightarrow \neg B$ would be also true with probability one, and hence $A \ \wedge \ B$ with probability zero. Yet, as a quantum proposition, the middle implication doesn't commute with the other two. Namely, the implications, equivalent respectively to
\[ \neg(A\ \wedge\ \neg S),\ \neg(S\ \wedge\ \neg R),\ \neg(R \ \wedge \ B),\]
are projectors to 3-dimensional subspaces orthogonal respectively to
\[ |00\ket-|10\ket, \ \ |01\ket, \ \ |10\ket-|11\ket .\]
The three subspaces intersect in a line (orthogonal to all the
three vectors) which is spanned (as one readily checks) by
\[ \psi:=\frac{1}{\sqrt{3}}\left(|00\ket+|10\ket+|11\ket\right) .\]
The proposition $A\ \wedge\ B$ is the rank-1 projector $AB=|\phi\ket\bra\phi|$, where 
\[ \phi=\frac{1}{2}\left(|0\ket-|1\ket\right)\otimes \left(|0\ket)-|1\ket\right) = \frac{1}{2}\left(|00\ket-|10\ket-|01\ket+|11\ket\right) .\] 
So, we find $\bra\psi|AB|\psi\ket=|\bra\psi|\phi\ket|^2=1/12 \neq 0$.
That is, quantum mechanics predicts that in a series of repeated measurements both $A$ and $B$ will be found equal to $1$ with non-zero frequency in contradiction with hidden variable theories.   
   
Combining this logic with variations of Wigner's friend setups (see the discussion of Bruckner's and Local Friendliness arguments in \cite{SYL}) one finds that the following two assumptions cannot be both true:

(i) an agent can apply unitary quantum mechanics and Born's rule to another agent, and

(ii) measurement outcomes are agent-independent (objective).

Another result of this sort is the Frauchiger--Renner theorem \cite{FR} which establishes that the assumption (i) cannot be true simultaneously with

(iii) an agent can use logical conclusions obtained by admitting the view of another agent. 

We show below that both types of conflicts are already visible in the much simpler original setup.

\medskip

{\bf 6. Wigner and his friend.} Recall that in this thought experiment Wigner's friend takes a measurement on a system S, while Wigner applies unitary evolution to obtain an entangled state, $\Psi$, of S+F and eventually takes a measurement on it to learn the outcome his friend observed. In the initial interpretation, Wigner and his friend disagree (as in the case with Schr\"odinger's cat) about the moment when the collapse occurs. But this ``contradiction'' is dissolved by noting that the two collapses happen in different systems. In later interpretations, the tension is not a contradiction but rather an ambiguity in the state assignment to S+F after the first measurement. If one is entitled to apply unitary evolution to an observer, then it should be the entangled pure state $\Psi$; but on the other hand, since that state contains the friend's knowledge of the measurement's outcome, it should be the mixed state, and from the friend's viewpoint --- even one of the collapsed states representing the outcome. How to find out which description is correct? Wigner's measuring $\Psi$ to find out the outcome (i.e. simply asking his friend or reading her lab journal) collapses the state --- whichever it was --- to the same one, and therefore does not answer the question. What Wigner can do to distinguish the two situations is measure on S+F an observable for which $\Psi$ is an eigenvector, e.g. measure the quantum proposition $W:=|\Psi\ket\bra\Psi|$ which yields 1 with probability one if S+F is in the pure state $\Psi$, and yields a statistical mixture of 1 and 0 otherwise. The problem with this plan is that the system F includes not only the lab journal, but the state of the friend's brain and the state of the environment, and so the measurement $W$ can only be made by an omnipotent version of Wigner --- that's why the whole dilemma remains merely a thought experiment.

For our further deconstruction of the problem, let us assume that S is a qubit, and note that although the lab and environment are described by enormous tensor products of Hilbert spaces, the only vectors that matter form a 2-dimensional subspace. It is spanned by $\Psi_1$ and $\Psi_0$ representing the states of F which correspond to the two outcomes of the first measurement, whereas $\Psi=|1\ket\otimes \Psi_1+|0\ket\otimes \Psi_0$. So, we may assume that F is also a qubit, and that the only unrealistic aspect of the situation is that it is an {\em intelligent qubit}: it can reason, apply quantum mechanics, and communicate with Wigner. 

Thus, let the initial state of the qubit S be $\psi=a |1\ket+b |0\ket$,
$|a|^2+|b|^2=1$, and so the entangled state of S+F is $\Psi=a |11\ket+b|00\ket$, where the left and right digits represent basis states of S and F respectively.

Note that if Wigner and his friend are observers external to the system S+F, then the situation is identical to the von Neumann model (with $Q$ renamed as $F$ --- we use italicized names of qubits to denote corresponding projectors). Measuring $S$ on $\psi$ is then interpreted as measuring $I\otimes F$ on $\Psi$. This projector does not commute with $W=|\Psi\ket\bra\Psi|$, and the results of the two measurements depend on the order. When $I\otimes F$ is measured first, the state $\Psi$ collapses to the mixture of $|11\ket$ and $|00\ket$ with probabilities $|a|^2$ and $|b|^2$. The subsequent measurement of $W$ yields $1$ with probability $|a|^2$ in the first case and $|b|^2$ in the second, i.e. totally with probability $|a|^4+|b|^4<1$. If, alternatively, $W$ is measured first, then $W=1$ with probability $1$, and the subsequent measurement of $I\otimes F$ yields $1$ and $0$ with probabilities $|a|^2$ and $|b|^2$ respectively. Note that $(I\otimes F=1)\ \wedge\ (W=1)$ is true with probability $|a|^4$ when $W$ is measured last and $|a|^2$ when $W$ is measured first: in quantum logic, naive Boolean expressions are ambiguous unless the corresponding projectors commute.      

Returning to the situation when the friend {\em is} the intelligent qubit F, we find, first of all, that assuming that the entangled state $\Psi$ of the system S+F ``knows'' the outcome of the first measurement is equivalent to introducing hidden variables and thus contradicts the well-tested principles of quantum mechanics. As we discussed earlier, the entangled state $\Psi$ is not the outcome of a measurement but merely the {\em a priori} theoretical description of what the state of S+F would be if the interaction between the system S and the apparatus F is done correctly so that their states are in perfect correlation with each other.

Next, being a part of the system, the intelligent qubit F cannot learn the outcome of the first measurement by measuring $I\otimes F$, can it? Wouldn't this be a self-referential act of the friend measuring her own mind (or, more generally, a system measuring itself)? We will see in Section 11 that this is exactly what happens in reality: human reflexive consciousness turns out to be an {\em external} observer of its own physical substrate. Yet for now, one can argue that the friend can measure $S\otimes I$ (which from her own perspective means she doesn't care that Wigner considers her a part of S+F and simply measures $S$). Note that in the von Neumann model this is not the measurement resulting in the entangled state $\Psi$, but its sequel whose outcome agrees with the first one with probability one. So, neither variant makes perfect sense, but just like $I\otimes F$, $S\otimes I$ doesn't commute with $W$, and the value of the ``proposition'' $(S\otimes I=1)\ \wedge\ (W=1)$ depends on the order in which the measurements are performed. The value is {\em true} with probability $|a|^2$ if $W$ is measured before the sequel measurement of S, and $|a|^4$ if after it. Thus, the ambiguity in the state assignment in Wigner's friend argument reflects the ambiguity of ordering non-commuting quantum propositions --- a problem, concealed by the agents' use of classical Boolean logic in their reasoning.

\medskip

{\bf 7. The Frauchiger--Renner argument.} Let us apply the method once more and rephrase the extension \cite{FR} of Wigner's friend thought experiment in terms of intelligent qubits. Following the description of it given in \cite{NR} and slightly simplifying their notation: there is a system R+S of two qubits prepared in the state $\frac{1}{\sqrt{3}}\left(|00\ket+|10\ket+|11\ket\right)$ on which Alice and Bob measure $A=R\otimes I$ and $B=I\otimes S$ respectively. Then Ursula and Wigner take certain binary measurements on the labs+environments of Alice and Bob respectively, and all four make some predictions, share some information, and compare some outcomes. So, we introduce into the system two intelligent qubits, A and B, whose states are entangled with R and S respectively according to the von Neumann measuring model for the measurements made by Alice and Bob. Thus, our total system has a $2^4$-dimensional Hilbert space whose basis vectors --- strings of four 0s and 1s  --- consist of basis vectors of four qubits A+R+S+B in this order. Overall the entangled state resulting from two measurement interactions becomes
\[ \Psi = \frac{1}{\sqrt{3}}|0000\ket+\frac{1}{\sqrt{3}}|1100\ket+\frac{1}{\sqrt{3}}|1111\ket.\]
In agreement with \cite{NR}, for the measurements performed by Ursula and
Wigner we can take orthogonal projectors to
\[ \phi:=\frac{1}{\sqrt{2}}\left(|00\ket-|11\ket\right) \]
in the spaces of two qubits: A+R for Ursula and S+B for Wigner, i.e.
\[ U:=|\phi\ket\bra\phi| \otimes I \otimes I, \ \ W:=I\otimes I \otimes |\phi\ket\bra\phi| .\]

The Frauchiger-Renner argument is entirely parallel to that in Bell's inequality. It involves three implications, (a),(b), and (c): the outcome $U=1$ implies $B=1$, which implies $A=1$, which implies $W=0$ (all with probability one) in conflict with (d): the chance that at the state $\Psi$ we have $(U=1)\ \wedge\ (W=1)$ is $1/12 \neq 0$. Implication (a) is obtained by noting that $B$ commutes with $U$, and that finding $B=0$ would collapse $\Psi$ to $\frac{1}{\sqrt{2}}\left(|00\ket+|11\ket\right)\otimes |00\ket$ where the first factor is orthogonal to $\phi$. Implication (b) is due to the fact that the common eigenvector $|01\ket$ of $R$ and $S$ with the eigenvalues $R=0$ and $S=1$ is orthogonal to $\psi$. Implication (c) is derived similarly to (a): $A$ commutes with $W$ and finding $A=1$ collapses $\Psi$ to $|11\ket\otimes \frac{1}{\sqrt{2}}\left(|00\ket+|11\ket\right)$ where the second factor is orthogonal to $\phi$. Finally, (d) is obtained by the same computation of $\bra \Psi|UW|\Psi\ket$ as before.

Note that the derivation of (a),(b), (c), (d) presupposes that the state $\Psi$ entangling the qubits R and S with their measuring qubits A and B already contains the outcomes of these measurements (as if the qubits by virtue of declaring them ``observers'' possessed the values of certain hidden variables). As we understand by now, this information should be gained by measuring $I\otimes R\otimes I\otimes I$ and $I\otimes I\otimes S\otimes I$ in the normal von Neumann model where Alice and Bob are external observers, or by sequel measurements of $A\otimes I\otimes I\otimes I$ and $I\otimes I\otimes I\otimes B$ if Alice and Bob impersonate our intelligent qubits. In either interpretation the former (Alice's) projector doesn't commute with $U$, and the latter (Bob's) doesn't commute with $W$.
Consequently, (a) is true only if Ursula's measurement precedes Alice's, (b) is true only if both Alice's and Bob's measurement precede Ursula's and Wigner's, (c) is true only if Wigner's measurement precedes Bob's, and (d) is true only if both Ursula's and Wigner's measurements precede Alice's and Bob's. In the case when the measurements of Alice and Bob come, as the protocol requires, before Ursula's and Wigner's, they collapse the state $\Psi$ to the equiprobable mixture of its three summands, each contributing $1/12$ --- as in (d) --- to the probability of $(U=1)\ \wedge\ (W=1)$, which therefore equals $3/12=1/4$.

Thus, the whole argument falls apart as long as one accepts the well-established fact of quantum mechanics that the values of entangled qubits don't exist until measured from the outside of the system. Let us note in parentheses that $a|11\ket+b|00\ket$ with $|a|^2=|b|^2=1$, considered as a tensor in $\CC^2\otimes \CC^2$, is a unitary isomorphism $\CC^2\to\overline{\CC}^2$ and is invariant under same unitary transformations of both planes. So, this entangled state doesn't even remember which observable's values it was meant to encode --- only that every observable of one qubit corresponds to a certain observable of the other.

To be honest, I am not even sure that our analysis undermines Frauchiger--Renner's intention. After all, they correctly claim that certain hypotheses are incompatible. Note, however, that the culprit here is the assumption that a conscious mind's reading the physical state of its own brain is not an additional external measurement. Interestingly, this view was spelled out in an early take on the measurement problem by F. London and E. Bauer which was recently revived \cite{Fre} by philosophers of the phenomenology school (see also \cite{WiBe}).  London and Bauer write \cite{LB} that the observer possesses

\begin{quote}
a characteristic and quite familiar faculty which we can call the ``faculty of introspection.'' He can keep track from moment to moment of his own state. By virtue of this ``immanent knowledge'' he attributes to himself the right to create his own objectivity --- that is, to cut the chain of statistical correlations summarized in $\sum \psi_k u_k(x)v_k(y)w_k(z)$ by declaring, ``I am in the state $w_k$'' or more simply, ``I see $G = g_k$ or even directly, ``$F = f_k$.''
\end{quote}

\noindent The sum here refers to the state of S+A+O where O is the observer, and in our terms the three statements are ``the outcome is $f_k$'', ``I see the apparatus pointing to $g_k$'', and ``I am in state $w_k$ of knowing the particular outcome.'' In retrospect, the last one reads like a slip of the tongue, as it contradicts the authors' own formal description of quantum mechanics where the state of a quantum system can only be measured (``introspected'' in this case) from the outside.

In any case, here the measurement paradox becomes inescapably intertwined with the mind--body problem. It is the focal point of the whole contention: advocates of the physicalist view would argue that an intelligent qubit is too small to carry its own value, but if a human conscious observer knowing Boolean outcomes of measurements cannot be described by quantum mechanics then the latter is not universal, for the outcomes are inscribed in the brain. This makes it imperative to bring some clarity to Chalmers' ``hard problem'',  as we will do in Section 11.  

\medskip

{\bf 8. The worldview.} The Frauchiger--Renner argument as well as all other extensions of Wigner's friend thought experiment are intended to test compatibility of various assumptions about quantum mechanics and thereby put constraints on the degree and meaning of ``realism'' that can be allowed by one or another {\em quantum interpretation}. However, our analysis points in a different direction. Namely, the common denominator of all those thought experiments is the assumption (i) that observers can be treated as quantum systems. We gave {\em a priori} reasons to expect that it breeds self-reference and hidden variables. Then we saw how both manifest in specific examples, and how the paradoxes come from tacit applications of Boolean logic to non-commuting quantum propositions. On the other hand, discarding the assumption dissolves all the paradoxes. Doesn't this mean that we should accept the inevitable and reject the assumption? This is what we get when we do. 

Quantum mechanics is a universal theory, but it does not describe the physical universe {\em per se}. Instead, it governs interaction between the object and the subject which are clearly separated from each other. The object is an isolated quantum mechanical system, and the subject is the community of observers of that system.  A state of the system (pure or mixed) is a device that to every quantum proposition associates the probability of it to be true in a series of identical ideal measurements performed by the community of observers external to the system. A measurement takes place when its outcome is recorded in a memory medium. Such measurements update (collapse) the state by the usual Bayesian rule of conditional probabilities. Between measurements, i.e. when the system remains isolated, the state evolves unitarily according to Heisenberg's (or Schr\"odinger's equation).

\medskip

Here is evidence that this picture is in good agreement with the views of classics (both quotes are from \cite{WiBe}, p. 19):

\begin{quote}

Most of us today feel that this necessary abandonment of a purely objective description of Nature is a profound change in the physical concept of the world. We feel it as a painful limitation of our right to truth and clarity, that our symbols and formulas and the pictures connected with them do not represent an object independent of the observer but only the relation of subject to object. But is this relation not basically the one true reality that we know? (E. Schr\"odinger)

\medskip

 \noindent [E]ven in science, the object of research is no longer nature but man’s investigation of nature. (W. Heisenberg)

\end{quote}  

By the way, the ``cut'' that according to Heisenberg is needed to delineate, somewhat arbitrarily, a quantum system from macroscopic classical measuring devices, is replaced with the split between the system on the one side and the observers with their memory devices on the other. This doesn't mean that one cannot develop physics of dissipative processes or apply science to humans, but that would still require memory devices and observers external to the studied object, and the latter cannot be used in a self-referential manner as its own recorder or trustworthy observer.

In order to apply Schr\"odinger's evolution one needs to choose the initial state of the system. Experimentalists know that {\em preparing} the initial state is essentially the same as measuring it. One can trace its unitarily evolution in reverse time, but the retrodiction obtained this way is counter-factual: the initial state was a result of the evolution-interrupting collapse that prepared it. Does the system have a state before anyone observes it? The following quote from P. Jordan seems to suggest that a negative answer makes more sense than the unverifiable and hence irrational belief that it does (quoted from \cite{Bell}):

\begin{quote} [O]bservations not only \emph{disturb} what has to be measured, they \emph{produce} it. In a measurement of position, for example, ... the electron is forced to a decision. We compel it \emph{to assume a definite position}; previously it was, in general, neither here nor there; it had not yet made its decision for a definite position ... If by another experiment the \emph{velocity} of the electron is being measured, this means: the electron is compelled to decide itself for some exactly defined value of the velocity ... we ourselves produce the results of measurement.
\end{quote}

\medskip

In summary, we come to a {\em dualistic} worldview, where Nature is akin to a {\em black box} which may have some hidden structure, but is not in any particular state until the universal observer interacts with it. The box responds randomly, at least as far as quantum mechanics is concerned, but there is sufficient regularity in this randomness to support rational prediction. 

On the surface, this worldview is at odds with the very successful paradigm of classical mechanics and science and general, according to which Nature functions the same way whether it is observed or not. Since classical mechanics is the limit of quantum mechanics, one still has to explain how the monistic classical world emerges in this limit from the dualistic one. In the rest of this paper we hope to answer this and some other pressing questions, and discover that in retrospect the dualistic paradigm seems valid rather tautologically and for reasons hardly related to quantum mechanics. 

Note that as a way to resolve the measurement paradox, the answer turns out to be purely philosophical: physicists of the ``shut-up-and-calculate'' school don't have to change anything in their theory and practice. Yet, the change in the worldview may shed some light on the way they should frame other problems. For example, recently L. Hausmann and R. Renner \cite{HR} pointed out a relation between Wigner's friend and the {\em fireball paradox} concerning black holes.

According to Hawking's theory, a quantum field in the presence of a black hole generates pairs of entangled particles of opposite energies, one of which falls into the black hole decreasing its mass, while the other escapes away as {\em Hawking radiation}. When enough particles fall in to dissolve the black hole, the radiated particles, paradoxically, are left entangled with nothing. (We realize that our description, based on the popularizing article \cite{Math}, might be only vaguely adequate, but we are not trying to make any technical point here.) 

The dualistic perspective points toward the solution known as {\em black hole complementarity}. The entanglement of two particles represents correlation between measurements of one and the other. But in Hawking's case, the community of observers outside the event horizon cannot measure anything on the particle inside the horizon, and even if they could safely send some members in to gain the access, they could not communicate with them across the horizon anyway. Thus, in this case the entanglement has no observable consequences to start with, and so it shouldn't have any at the end. All the outside observer sees is the black hole slowly evaporating through the Hawking radiation.


\section*{Anthropology}

{\bf 9. Civilization, practice, culture.} We use these terms in parallel with {\em individual, behavior, consciousness}, i.e. referring to respectively objective (practice, behavior) and subjective (culture, consciousness) components of the whole (civilization, individual). We need these notions because the principle of inter-subjectivity of observation makes the entire human civilization the ultimate candidate on the role of {\em the universal observer}. One can talk of {\em a} civilization not necessarily in anticipation of contacts with space aliens, but in the common sense of distinguishing the Celtic civilization from the Roman one, though the latter eventually absorbed the former. Also, it is clear to us that {\em Stonehenge} didn't assemble itself --- some prehistoric humans labored to transport the monoliths and erect the peculiar edifice. But the significance of it for them escapes our understanding because we are ignorant of their culture. This illustrates the meaning in which ``culture'' is subjective relative to the civilization carrying it, while practices (e.g., of moving stones) are potentially observable even by ignoramuses and hence objective.

Humans are weak individually but strong as a group; so, we are social animals, though not the only ones. Apparently, many if not all social animals develop language: it is needed for coordinating group activities. Yet, as far as we can tell, we seem to be the only one who evolved a culture. To understand better what that is, let us try to imagine how it could emerge (and leave it to archaeologists and linguists to decide whether the concepts we introduce are useful).

Imagine a prehistoric group of people hunting mammoths by digging a pit and chasing the animal toward it. Their language contains pointers to specific things and actions: {\em mammoth, pit, chase}, whose shared meaning is acquired by the group's members from the shared practice of hunting.

Altogether these pointers are arranged into what we call a {\em plan} of hunting, but the group doesn't need to know they are executing a plan: they just execute it. Yet, from several specific plans --- of hunting, fishing, building --- the concept of {\em a plan} can be abstracted. This is a pointer to an {\em ideal} object. Once acquired, the idea of ``plan'' has a creative capacity: the group can come up with a plan of doing something else. This is a germ of culture.

From several such secondary pointers --- not to specific things or actions, but to ideal objects: {\em plan, goal, success} --- next-level pointers can be abstracted: such as e.g. {\em idea, purpose}, and so on.  Let's define a culture as such a layered web of abstractions which can point not only to directly observable objects and actions, but also to ideal things: other pointers, pointers to pointers, etc., including e.g. {\em mind, concept, theory}. Going far beyond a simple language, this structure helps to evolve --- and is evolved within --- a civilization. The latter does not merely reproduce the same patterns of group behavior, but is capable of improving its practices and adapting its environment in a purposeful and creative fashion.

\medskip

{\bf 10. The black box.} The next thought, which seems to me true tautologically, but which is usually hard to get across, is that the world does not intrinsically consist of individual objects we believe it consists of: it is we who designate them as such and divide the world into specific ingredients with our concepts.

What is a pine-tree in my back yard? Is is a live plant producing pine cones, or a photosynthesis laboratory, or potential firewood, or --- for the insects who live in it --- their habitat, or a population of individual cells with interrelated metabolism, or --- according to the R. Dawkins' concept of selfish genes --- a means for their proliferation, or a conglomerate of atoms, or essentially the empty space between the atoms' nuclei, or literally the empty space if you are a neutrino, or --- if you are made of dark matter --- a negligible bump on the fabric of space-time?

It is tempting to say that the tree is all of the above, that the concepts we introduce are not arbitrary but reflect the objective reality, and that any civilization would come up with an equivalent system of concepts. The last point is verifiably correct for human civilizations on Earth, but this is because their practices are very similar. Yet, this would hardly remain true for a civilization of space aliens made of dark matter (and adding ``should we encounter one'' is meaningless: for us, they are not encounterable).

So, the point is that the concepts we introduce reflect not the reality {\em per se} but our adaptive {\em practice}. In order to purposefully and successfully modify our environment, we need to foresee consequences of our actions based on the memory of our past experience even when that experience does not include that particular type of action. When we succeed, we call it {\em cause and effect}. One can debate whether there is a deeper, culture-invariant, objective source of causality, but our subjective time arrow is the most direct one.

Thus, we divide the black box of our environment into its constituent parts with our concepts, ask how the parts interact with us or each other, memorize the results, and try to predict future outcomes. What we call ``objective reality'' is nothing more than the history of such interaction between the universal observer (our civilization) and the black box. Note that by saying this I merely paraphrase Schr\"odinger's ``But is this relation not basically the one true reality that we know?''

This dualistic coexistence on an equal footing of the object and the subject derived from our current anthropological standpoint fits well with the lessons of the measurement problem in quantum mechanics. Both imply that {\em without us}
the black box is not in any particular state --- it is {\em featureless}. This doesn't mean that without us the world disappears, but it means that nothing specific happens in it. For anything specific to happen, someone needs to conceptually divide the ``soup'' of everything into its ingredients and ask how they interact. And without us there seems to be no one available to do the dividing.

The usual and immediate objection to this is that clearly dinosaurs existed before humans. Sure, many things happened {\em before us}, but this is not the same as {\em without us}, because tautologically their very happening --- within the one true reality we know --- is a result of us looking back in time and applying our current conceptual framework. Still, this refutation sounds like a linguistic stunt, because there are good empirical reasons to insist that dinosaur fossils were {\em actually} lying there regardless of whether we knew about them or not. This is an important point, and we will later discuss these reasons and how they square with quantum mechanics, which essentially says that things don't exist until observed.

Having explained in what sense the black box is ``featureless'' until it is endowed with specific features by observation, we should stress that this doesn't necessarily mean it is ``structureless.'' We successfully model many phenomena by differential equations. Still, this can be understood as the observer's imposing its conceptual framework on the universe: focusing on the processes and quantities in physics, chemistry, ecology which have well-defined rates of change. However, the discovery that the dynamics of isolated conservative mechanical systems is {\em Hamiltonian} cannot be dismissed as an observer-induced artifact. Specific Hamiltonians are a result of choosing a system of study and measuring its parameters, but the very property of being Hamiltonian (which mathematically requires a peculiar geometric structure of the phase space) says something about the intrinsic structure of the black box --- as does the formalism of quantum mechanics.   

\medskip

{\bf 11. The hard problem.} Accepting the dualistic worldview or not, one faces this problem, because both the culture-endowed civilization and the community of observers in quantum mechanics supposedly consist of conscious individual agents. So, what is consciousness? Or more concretely: how do physical brain processes give rise to the 1st-person subjective experiences?

To this hard question, there is an easy answer: they don't.

First, the division into objective and subjective, as any conceptual division, is done for the benefits of using the concepts in a discourse. It is not to separate two non-overlapping realities, but to assess a single one from two different angles --- from the 3rd-person and 1st-person perspectives in this case.

From the 3rd-person position, there is a continuous spectrum of physical processes beginning with simple feedback loops and ``behaviors'' of subway turnstiles, to increasingly complicated nervous systems and intelligent or even creative behaviors found in the animal kingdom, to AI-assisted robots and LLMs matching or even exceeding human brains in their processing capabilities. Understanding the functioning of a brain is indeed a hard and exciting scientific problem, and neuroscientists are making spectacular progress in this domain.

However, we are raised with the understanding that while experiencing, say, pain --- which manifests in the objectively measurable reactions of our hormonal system and neurons ---  we also have the subjective, 1st-person assessment of it, and can say (or merely think),  ``Oh, my God, I am in pain!'' Discussing such reactions is the discourse where the concept of subjectivity comes handy.

Note that in this example, the subjective reaction to pain uses language and is therefore culture-induced or at least culture-assisted. The hard problem of consciousness comes from the belief that there is an intermediate layer between the objective physical reaction of the brain and body, and the subjective but culture-induced human reflexive consciousness: something innate, pre-cultural, present in sufficiently developed animals, and yet subjective, accessible only to the experiencer. Of course, in regard to animals, this is merely a projection: we observe behaviors (i.e. objective manifestations) which we attribute to that layer of subjectivity in ourselves. Thus, the question reduces to this: How do we know that we have this pre-reflexive, animal-like layer of subjectivity?

Being subjective, it cannot be measured or confirmed through behavior. So, the only available answer is: ``I feel it, I just know it is there!''  The problem with this answer is that it itself is an act of human culture-assisted reflection. In other words, this hypothetical layer of consciousness --- allegedly intermediate between the objective physical processes and human culture-induced subjectivity --- cannot manifest in any way except the other two. It is merely a belief, and the grounds for it are just as rational as in ``I feel Jesus in my heart!''

Now Occam's razor helps us dispose of this figment alongside the ``hard problem.'' Our 1st-person subjective consciousness turns out to be not a product of our brain, but a slice of human culture downloaded into it (or rather trained --- not unlike an LLM into its data center) through upbringing and practice.    

I realize that this understanding that human reflexive consciousness is not a product of brain is deeply foreign to the Western tradition focused on ``qualia.'' Let me note however, that it is quite natural and wouldn't sound original in, say, Marxist tradition. As an illustration: Lev Vygotsky (1896--1934), the Soviet {\em vis-\`a-vis} of Jean Piaget, formulates an empirically motivated conjecture \cite{Vyg} that egocentric (not directed at anyone) speech of a child is a developmental precursor of the internal monologue (and not the other way around, as it was assumed by Piaget).  

In his book \cite{Aha}, R. Aharoni argues that all fundamental puzzles of philosophy come from confusion between the 1st- and 3rd-person perspectives. In particular, he posits, our 1st-person experiences consist of our thoughts about the state of our body and mind examined from a 3rd-person position as with any other external object. Here is a metaphor inspired by that explanation.

I have a view from my window, you have a view from yours. I don't have direct access to your view and you to mine --- in this sense our views are subjective. You can describe your view to me, but it is not the same as me actually seeing it. However, if I could see it, I would examine it the same way I am examining mine: the only difference is in the degree of access.   

The same is with our 1st-person experiences. My enculturated human reflexive self examines the physical state of my body and the memory of my thoughts the same way as it would examine the state of yours should I be given access to them. The difference is that I have many more channels of information about mine than I have about yours. We sould add, though, that in practice, the borderline between my subjective culture-induced assessment and the physiological functioning of the brain might be ephemeral --- as is every conceptual division. (Even the seemingly black-and-white distinction between genders looks up close more like a rainbow.)

We'd like to stress, however, that again our purely anthropological take on the problem intertwines with the principles and paradoxes of quantum mechanics. The point argued above is that awareness of the state of our own mind is an act of external observation. As we noted in Section 7, assuming the opposite lies at the heart of all Wigner's friend-like paradoxes and goes back to London--Bauer's ``introspection'' passage where an agent's mind considered as part of a conservative quantum mechanical system is still expected to access its own state without any external measurement.

Once we realize that an observer's introspection is enabled not by the mysterious internal connection between her brain and mind, but by her place within the universal observer as the carrier of her private slice of inter-personal culture, the paradoxes resolve. For the introspection becomes a special instance of {\em external} measurement by the universal observer over the black box of its environment.     

\medskip

{\bf 12. Epistemic = ontic.} Having demystified consciousness, we can now pinpoint the ontological status of wave functions.

On the surface, a quantum state, pure or mixed, has only epistemic significance: it characterizes the observer's reliable prediction of expectation values in an ensemble of identical measurements of every observable. An act of measurement is an objective, material interaction of the formerly isolated system S with a measuring apparatus A. In this theoretical description of an idealized measurement, there is no need to split the process (as we did in the von Neumann model) into unitary entanglement between S and A followed by measuring the entangled state --- it is enough to posit that the memory-equipped apparatus records the outcome for posterity. In practice the ``posterity'' may last milliseconds after which the outcome in a computer's memory is overwritten, but this is irrelevant in principle: it suffices to know that the interaction and recording occurred, and the outcome was, or at least could be made, available to the observer. In either case the state of the system (i.e. the state of knowledge of the observer about expectation values of further measurements conditioned on this outcome) is updated, and the projection rule expresses the ``collapse'' of the pre-measurement state into the updated one.

It is useful to compare this abstract description with a real-life procedure, such as e.g. the Hitachi double-slit experiment \cite{Hit} where electrons are shot one-by-one through a magnetic lens which forces them into two different paths. The locations on the screen, where the electrons are absorbed, and which eventually assemble into an interference pattern, are recorded in a movie which is then posted on YouTube. So, the answer to the usual provocative question {\em Whose consciousness collapses the wave function?} depends on whose knowledge the wave function is meant to represent. If you are the kind of person who needs to see for yourself before you believe it, then the wave function characterizes the state of your personal knowledge, and it collapses only when you watch the movie. However, as we have discussed, states of quantum systems are inter-subjective, and this one actually characterizes the state of knowledge of the team of scientists and engineers who ran the experiment. Furthermore, once posted on YouTube, the measurements become accessible to our entire civilization. In this sense, the wave function and its collapses represent the state and change of knowledge of the universal observer. Thus, the real (as opposed to idealized) observation is a complex social activity which employs sophisticated validation mechanisms (after all, the movie was easy to produce without any experiment) involving certification of the scientists and engineers, personal and corporate prestige, journal refereeing and publication, and so on.   

Our point, however, is that since ``the only true reality that we know'' is the history of interaction between the universal observer and the black box of its environment, the purely epistemic status of quantum states becomes indistinguishable from their ontic status as {\em the} states of that objective reality.      

By the way, this is one of several distinctions of our dualistic worldview from the ``relational'' interpretation of quantum states by C. Rovelli \cite{Rov}, who denies them any ontic value though also views them as epistemic (or information-theoretic) relations between a system and its observer. (The reductionist attitude which, in particular, allows him to interchange the roles of the system and observer, is another.)       

\medskip

{\bf 13. Esse est percipi.} This famous Latin maxim by 18th century philosopher George Berkeley summarizes what's often considered the quintessence of ``subjective idealism'' --- and coincidentally of quantum mechanics: {\em To exist means to be observed}. Still, Berkeley couldn't deny things their objective existence, and for a good reason: together with (as Piaget explains) every toddler, Berkeley knew that a ball that rolls out of the view under a table doesn't cease to exist and can be found there. So, he employed God as the universal observer who sees everything and thereby grants the world its existence. Thus, the worldview we derived from the measurement paradox in quantum mechanics is essentially a revival of Berkeley's views, where, however, {\em God} has to be replaced with {\em Culture}.

Yet, culture is not as eternal as God, and so we need to examine those empirical reasons which normally convince us that things can exist even if never observed. 

Following a prompt from R. Penrose \cite{Pen}, imagine a newly discovered planet in Alpha Centauri. Our knowledge of its precise position is uniformly distributed along its circular orbit. Suppose two space agencies independently send their probes and measure the position more precisely. Our experience says unambiguously that the two independent measurements will not be randomly selected from the uniform distribution, but will be in perfect correlation with each other. The simplest explanation of this correlation is that the planet was {\em actually} located at that position, and it is only our knowledge of it that was incomplete. Moreover, if between the two measurements, unknowingly to those two agencies, a third one measures the momentum of the planet, this won't affect the position and the correlation.

Note that the same experiment with a quantum harmonic oscillator yields different results. Two consecutive measurements of position $q$ will produce the same outcome if no measurement of momentum $p$ occurs in between, but become uncorrelated otherwise. Since planets and toy balls are objects just as quantum as harmonic oscillators, fundamentally speaking, our classical intuition based on celestial observations and domestic experience is incorrect. It works only because in the quantum commutation relation
\[  pq-qp =i\hbar \]
the Planck constant $\hbar \approx 10^{-34}\ kg\cdot m^2\cdot s^{-1}$ is vanishingly small by our macroscopic standards. Consequently, in the classical limit the observables practically commute and so their values can be measured simultaneously.

As an issue tangential to our present discussion, at this point one is expected to bring up {\em decoherence} which supposedly explains why macroscopic objects are not found in quantum superposition states. The idea is that no real system is perfectly isolated, and so its state is constantly ``monitored'' by interactions with the environment (the random bombardment by cosmic rays, or --- according to the famous remark by M. Berry \cite{Ber} --- even by the gravitational effects of particles from the outskirts of the visible universe) which kill interference and allegedly collapse the state to the classical trajectory of the object.

While the phenomenon of decoherence is certainly real and presents genuine difficulties in quantum computing, I am not sure that it is relevant to explaining the observed classicality of the macroscopic world. On the one hand, it looks suspicious (especially amidst the discussions of {\em space} and {\em time} being ``emergent'' phenomena) that the interference terms (i.e. off-diagonal entries of the density matrix) are killed specifically in the basis of positions, or momenta, or other culturally-meaningful observables (the so-called {\em preferred basis} problem). On the other hand, the commutativity of observables in the classical limit seems sufficient.

Indeed, what does it mean to detect a toy ball in the superposition state $\Psi$ of being {\em here} and {\em there}? For this, as in Wigner's friend experiment, one should measure the observable $W:=|\Psi\ket\bra\Psi|$ for which $\Psi$ is an eigenvector. One obvious obstruction to even guessing such an observable is the ambiguity of {\em here} and {\em there}. Fortunately, measuring is the same as preparing a state. It suffices to choose {\em here = in my hand}, {\em there = under the table}, and perform an experiment that measures $W$ on the state {\em here}: this would put the ball in superposition of being {\em here} and {\em there}. But no one, it seems, has come up with a {\em realistic} experimental scheme that, even neglecting decoherence, would accomplish this with a macroscopic object like a toy ball. An informal explanation of this may look like this. Classically, if the statistical mechanical density of the ball is the delta-function concentrated at one phase point, then it will remain concentrated at a single phase point during evolution under any Hamiltonian $H(p,q)$. For the quantized Hamiltonian $\hat{H}$, the commutators $[\hat{H},\hat{q}], [\hat{H},\hat{q}]$ are of order $O(\hbar)$, and hence semi-classically the narrow wave packet representing the initial state will remain narrow during the long Ehrenfest time of a macroscopic object. (It would more be useful to have a formal estimate here under the assumption that the Poisson brackets $\{ H, p\}, \{ H, q\}$ are of order $O(1)$.) 

Thus, what really happens, is that before any observation of the ball its state is ambiguous, i.e. described by a statistical mixture of many pure states. But once it is observed to be {\em here} (i.e. with definite modulo $\Delta p \cdot \Delta q \approx\hbar$ position and momentum), no realistic action can place it in the superposition.    

Returning to the empirical grounds for our classical intuition, let us note that besides ideal quantum measurements there exist events so destructive to the unitary evolution of an isolated system that they affect even macroscopic objects: the toddler's mom could secretly take the ball from under the table and put it in the toy box. The objectivist point of view on this possibility is that one needs to include the mom into the system and describe it by laws of physics. As we discussed earlier, this does not lead to Wigner-like logical contradictions in classical mechanics because classical observable commute, but fundamentally this worldview is unsustainable, as it ultimately makes observers conduct a self-referential study of a system they are part of. So, we should understand that at least in domestic affairs, the observers themselves interrupt the reversible evolution of the classical or quantum black box by e.g. staging their experiments and otherwise interacting with it. Still, let's take another look at cosmic events where observers' intrusion is irrelevant.

Suppose we have determined the position and momentum of the planet circling Alpha Centauri. Then Kepler's laws allow us to determine the future {\em as well as the past} whereabouts of the planet. Doesn't this mean that those whereabouts exist regardless of our knowledge of them? We have an overwhelming amount of experience validating Kepler's laws to view these prediction and retrodiction perfectly reasonable. Except that we should realize that they are {\em conjectural}: the planet doesn't have to be found at the predicted location nor had it to be in the retrodicted one. It is possible that the current state of the planet, or even the planet itself, are a result of a catastrophic collision that happened in the recent past. Of course, given what astronomers know about such events, they are very rare, and if it occurred relatively recently, there would be remnants of that collision flying around. However, if order to exclude the possibility, additional observations are required together with more complicated several-body models.

In other words, we should not forget that all our models are idealizations. We single out a relatively small part of the black box as a reasonably isolated system, and apply our theory to it, but the assumption that the entire black box is described by a similar model is an invalid extrapolation. It is up to the universal observer to decide what system to consider, and which measurements to make, whereas the idea, that the black box includes self-appointed planets (exhibit A: Pluto) with well-defined trajectories existing outside the knowledge state of the observer, has to be abandoned.

By the way, once we realize that concepts of classical mechanics, including mathematically well-defined phase points, are merely idealizations --- e.g. because of quantum uncertainties, or the Planck scale of $10^{-35}\ m$ below which, physicists say, the space continuum itself becomes meaningless --- we should conclude that the determinism of classical trajectories is also an idealization. When exponential divergence of mathematically well-defined phase trajectories is coupled with, say, Planck scale physical uncertainty of the system's parameters and/or initial conditions, it turns out \cite{GivI} that billiard trajectories can become ill-defined after several dozen of reflections, the trajectories of gas molecules in the solid ball model live only nanoseconds, and the trajectories of comets and planets can lose meaning well within the lifespan of the system. Furthermore, this observation undermines the notion of {\em definite past}  even in the classically construed physical reality.

Perhaps, determinism of the Schr\"odinger evolution is also an idealization, but let's not intrude into the territory of quantum gravity.

\medskip

{\bf 14. Dinosaur fossils.} We often referred to irreversible memory devices as irreducible constituents of the cognitive process. But why do they exist at all?

One can make a memory bit by letting (or not) gas molecules, forced to the left half of a chamber, spread to the entire chamber. Actually the Poincar\'e recurrence theorem guarantees with probability one that one day all molecules will return to the left half. The catch is that the age of the universe is expressed by an 18-digit number of seconds, while the number of digits expressing the Poincar\'e recurrence time is 23-digit itself. It is not the first principles of conservative dynamics, but this grotesque discrepancy in time frames what accounts for irreversibility of our clumsy memory device.

A more practical computer bit can be classically modeled in terms of a double-well potential: a particle sinks into one of the wells and dissipates its kinetic energy through friction. A quantum-mechanical description of this involves two stationary states of the particle: the ground state $\psi_0$ and the excited state $\psi_1$, while states localized in one of the wells are represented by their linear combinations, say: $\psi_{\pm}:= (\psi_0\pm \psi_1)/\sqrt{2}$, and are not stationary. Eventually $\psi_{\pm}$ are guaranteed to evolve into $\psi_{\mp}$, but the time needed for this is astronomically long if the barrier between the wells is big enough. On the other hand, in order to record a particular value of the bit  one needs to temporarily tilt the potential toward one of the wells. If the goal is measuring a quantum proposition $P$ and recording the outcome, this tilting can be modeled by adding to the Hamiltonian $H_{system}\otimes I+I\otimes H_{bit}$ the interaction term $h(t) \left(P-\frac{1}{2}\right)\otimes W$, where $W$ is the tilting potential, and $h(t)=1$ during the tilting time interval and 0 outside it. At the end of the recording, the excess kinetic energy created by the tilting is absorbed by the environment. Thus, a unitary description of the event becomes equivalent to the formal scheme of Section 3 which produces the entangled state
\[ |\Psi\ket = |+\ket \otimes |\psi_{+}\ket \otimes |\phi_{+}\ket \ +\ |-\ket \otimes |\psi_{-}\ket \otimes |\phi_{-}\ket.\]
Here $\psi_{\pm}$ are the idealized versions of the (actually time-dependent) states of the recorder ``trapped'' inside respective potential wells for a long time, and $\phi_{\pm}$ are the corresponding states of the uncontrollable environment.

Thus, by tracing out the environment, the bit is {\em a priori} found in a mixture of its two values, but to recover the stored value an outside observer still has to measure $I\otimes Q\otimes \tilde{I}$ on this system and record the outcome in a memory medium external to the system. So, the role of memory devices remains fundamentally irreducible, although they can be described by physics. And as we see, their irreversibility is not derived from first principles of unitary evolution, but is due to the relative stability ({\em meta-stability}) of certain states on culturally relevant (or even far longer) time scales.

Closer to fossils, consider the situation typical for modern physics experiments: a computer storing data bits for years before they are processed and acknowledged by the researchers. Certainly, this is the moment when the data enter (collapse) the {\em state of knowledge} of the universal observer. But did the values of the bits exist before? To answer, imagine someone else reading the value a year earlier; then due to the meta-stability of the states $\psi_{\pm}$ (and ignoring in this theoretical discussion the possibility of memory corruption) the value of the bit observed now would be the same. In this sense, it is safe to assume that the collapsed state existed since the moment of recording (and running Schr\"odinger's evolution of it in reverse time confirms this).

With dinosaur fossils, it is similar. Nature itself creates memory devices. Once a fossil is discovered, this collapses the knowledge state of the universal observer. Yet, imagining that someone examining the same site a century earlier found it fossilless is (again, modulo possibilities of scams, etc. irrelevant for our theoretical discussion) in conflict with the meta-stability of the states $\psi_{\pm}$ under reverse-time Schr\"odinger's evolution.

Schr\"odinger's cats can be analyzed in similar terms, but we leave this to the reader. 


\medskip

{\bf Acknowledgments.} I am thankful to Anton Kapustin for numerous discussions of various aspects of the measurement problem, to Ron Aharoni for sharing a draft of his philosophy book, and to Gemini and especially ChatGPT for many productive brainstorming sessions.

\enddocument